# Day–night cloud asymmetry prevents early oceans on Venus but not on Earth

Martin Turbet[1], Emeline Bolmont[1], Guillaume Chaverot[1], David Ehrenreich[1], Jérémy Leconte[2], Emmanuel Marcq[3]

**Earth has had oceans for nearly four billion years[1] and Mars had lakes and rivers 3.5-3.8 billion years ago[2]. However, it is still unknown whether water has ever condensed on the surface of Venus[3,4] because the planet – now completely dry[5] – has undergone global resurfacing events that obscure most of its history[6,7]. The conditions required for water to have initially condensed on the surface of Solar System terrestrial planets are highly uncertain, as they have so far only been studied with one-dimensional numerical climate models[3] that cannot account for the effects of atmospheric circulation and clouds, which are key climate stabilizers. Here we show using three-dimensional global climate model simulations of early Venus and Earth that water clouds – which preferentially form on the nightside, owing to the strong subsolar water vapour absorption – have a strong net warming effect that inhibits surface water condensation even at modest insolations (down to 325 W/m$^2$, that is, 0.95 times the Earth solar constant). This shows that water never condensed and that, consequently, oceans never formed on the surface of Venus. Furthermore, this shows that the formation of Earth's oceans required much lower insolation than today, which was made possible by the faint young Sun. This also implies the existence of another stability state for present-day Earth: the 'Steam Earth', with all the water from the oceans evaporated into the atmosphere.**

[1]Observatoire astronomique de l'Université de Genève, Chemin Pegasi 51, 1290 Versoix, Switzerland. [2]Laboratoire d'astrophysique de Bordeaux, Université de Bordeaux, CNRS, B18N, Allée Geoffroy Saint-Hilaire, F-33615 Pessac, France. [3]LATMOS/IPSL, UVSQ, Université Paris-Saclay, Sorbonne Université, CNRS, Guyancourt, France.

Most numerical studies that seek to identify the conditions that allow terrestrial planets to have surface liquid water (in the form of lakes, seas or oceans) assume that surface liquid water was present in the first place. Studies of Mars have focused on finding the conditions necessary to prevent complete glaciation[8], in an attempt to explain widespread evidence of past intermittent hydrological activities[2,9,10]. Studies of Venus have focused on finding the conditions necessary to delay complete evaporation[4,11]. Deuterium-hydrogen (D/H) isotopic ratio measurements[12,13] (about $10^2$ times higher than on Earth) suggest indeed that the early Venus superficial water reservoir ranged from roughly 4 to 500 m global equivalent layer[4] (GEL; that is, the thickness of the liquid water layer if spread evenly across the surface), but could have been even higher as the D/H fractionation factor during (hydrodynamic) escape of hydrogen can be close to unity[14]. The early Venus superficial water reservoir was thus much higher than today[5] (about 3 cm GEL) and was possibly in the form of a liquid water ocean[4,11,15,16]. Studies of Earth have focused on finding the conditions necessary to prevent both complete glaciation[17,18] and evaporation[19,20]. This is not only to match the geological record and solve the so-called faint young Sun paradox[1,17,20,21], but also to predict the future of Earth[19,20]. These studies have demonstrated that inherently three-dimensional (3D) atmospheric processes are key stabilizers for maintaining a surface liquid water ocean. Of particular interest for this study, the slow rotation of Venus could lead to the formation of subsolar clouds, which would effectively reflect the incident solar flux and thus stabilize oceans[4] (Fig. 1a). On fast-rotating planets like Earth, atmospheric dynamics produce dry regions (in the descending branches of atmospheric cells, for example, Hadley cells) that would increase thermal emission to space, cooling and thus stabilizing oceans[19,22].

However, even before the question of the conditions for maintaining a surface liquid water ocean arises, water initially present in the young and warm planetary atmosphere must be able to condense on the surface. Planets are indeed expected to form hot due to their initial accretion energy, and thus to cross a magma ocean stage[3,16,23,24] – where superficial water is present only in the form of vapour – before evolving towards their final state. The conditions leading to the condensation of a water ocean after the magma ocean phase have so far been studied with only one-dimensional (1D) numerical climate models[3,24], which neglect the effects of atmospheric dynamics and clouds. Hot, water-vapour dominated atmospheres are indeed notoriously difficult[19] to simulate in three dimensions because they require an adaptation of condensation, convection, radiation and convergence schemes in a regime where water vapour can condense and yet be the dominant gas.

Here we perform 3D global climate model (GCM) simulations designed to simulate the set of conditions required for water condensation and, consequently, ocean formation to take place on terrestrial planets, and in particular on early Venus and Earth. For this, we adapted the GCM to cope with the extreme environments of hot, water-dominated primitive atmospheres, as well as their extremely long convergence time (see Methods). Simulations of initially hot (initial temperature around $10^3$ K) and water vapour-dominated ($H_2O$ partial pressure 10 bar) atmospheres of early Venus and Earth (forced to insolations between 300 and 675 W/m$^2$, defined here as average top-of-the-atmosphere fluxes) reveal that clouds are essentially concentrated on the nightside (Fig. 2a, b, e, f), independently of the planetary rotation speed and day duration. The absence of clouds and strong solar absorption by water vapour on the dayside reduce the planetary albedo. The presence of clouds on the nightside produces a strong greenhouse warming, which effectively reduces the nightside thermal cooling to space (Fig. 2c, d, g, h). As a result, clouds have a strong net warming effect (Extended Data Fig. 2).

Preferential nightside water cloud formation is mainly driven by the direct visible and near-infrared absorption by water vapour – here the dominant gas – of the incoming solar radiation. First, the absorption by water vapour produces a strong atmospheric heating in the subsolar region – which has already been qualitatively observed[25,26] – that efficiently warms the atmosphere down to about 1 bar (Fig. 3b), which breaks moist convection, and therefore strongly inhibits dayside cloud formation. The dayside cloud cover in fact decreases with increasing incident flux, as illustrated qualitatively in Fig. 2a, b, e, f and quantitatively in Extended Data Fig. 3b. For high insolation (that is, about 650 W/m$^2$, which is equivalent to the solar flux on present-day Venus), the dayside is almost entirely depleted in clouds, and the bond albedo reaches closely that of previous state-of-the-art 1D numerical climate model cloud-free calculations[27–29] (Methods). Figure 3d in fact shows that this direct subsolar absorption is up to about $10^4$ stronger than in standard 3D GCM simulations where all water is initially assumed to be condensed on the surface (labelled 'cold start' in Fig. 3). This is due to the fact that the water vapour mixing ratio is up to about $10^7$ larger, at least in the upper part of the atmosphere (Fig. 3c).

In our baseline simulations of hot and steamy Earth and Venus, the atmospheric circulation is strongest in the upper atmosphere (above about 0.1 bar). It is dominated by a strong eastward jet that is even super-rotating in the early Venus simulations, owing to the slower planet rotation. It is also characterized by a Brewer-Dobson-like circulation (Fig. 3a) which transports warm stratospheric air parcels from the equator to the poles. As the warm and dry stratospheric equatorial air parcels get transported eastward and poleward by the winds, they cool by thermal infrared emission, which eventually leads to preferential cloud formation on the nightside, specifically at two gyres, and at the poles (Fig. 2a, b, e, f), where the balance between thermal infrared cooling and dynamical heat redistribution heating (by the winds) is favourable. The mechanism leading to the preferential formation of clouds in the nightside is depicted in Fig. 1b. The accumulation of clouds on the nightside reduces the thermal cooling to space. Figure 2c, d, g, h shows that the thermal emission to space is actually anticorrelated with the position of water clouds, which produce a strong greenhouse effect. This peculiar distribution of clouds produces a strong net warming effect, between 50 and 120 W/m$^2$ in the range of insolation of around 340-675 W/m$^2$, that is, that of Earth and Venus (Extended Data Fig. 2). At high insolations, nightside temperatures increase with insolation, resulting in a reduction in cloud cover and thus in greenhouse warming of clouds.

The nightside cloud pattern and its net warming effect are robust to a wide range of insolations (Fig. 2). They are also robust to a wide range of cloud properties, total amount of water vapour and addition of carbon dioxide ($CO_2$) (Methods). The preferential formation of stratospheric nightside water clouds in both the Earth (sidereal rotation period $P_{rot}$ ~ 24 h) and early Venus simulations ($P_{rot}$ ~ 5,833 h, that is, about 244 times that of Earth) suggests that the mechanism is also robust to a wide range of rotation periods. A comprehensive sensitivity study is needed to quantitatively confirm this result, as it has been shown that cloud and atmospheric circulation feedbacks can vary nonlinearly and non-monotonically with rotation period[30]. The mechanism differs from the stabilizing tropospheric subsolar cloud feedback[4,31] (Fig. 1a), which works only for slowly-rotating planets[31-33]. This is because the stratosphere adjusts here quickly to the solar heating perturbation (constantly destroying clouds on the dayside and reforming them on the nightside) compared with the troposphere, which requires a much longer daylight period to establish a region of stable and intense moist convection, capable of producing thick, reflective clouds[32].

Cloud formation in the GCM is largely dominated by large-scale condensation (that is, water vapour condensation driven by large-scale air movements, as illustrated in Fig. 3a) and not by subgrid-scale condensation (driven by small convective cells, as in refs[4,31]). This indicates that our results should be weakly sensitive to the choice of the subgrid moist convection parameterization, which is a strong indication[34] that preferential nightside cloud formation is likely to be a robust mechanism across GCMs.

As a direct consequence, the net warming effect of clouds markedly reduces the insolation required to condense water on the surface, compared with previous 1D cloud-free calculations[3,16,24]. The simulations we have carried out at multiple insolations reveal that (1) the insolation required to condense water on early Venus is about 325 W/m$^2$ (or about 0.95 times the Earth solar constant) and (2) that on Earth is about 312.5 W/m$^2$ (or about 0.92 times the Earth solar constant). When the initially hot and steamy planetary atmospheres reach these insolation thresholds, clouds start to form on the dayside, which makes the bond albedo suddenly jump, and produce a net cooling of the atmosphere until water vapour starts to condense on the surface, forming oceans (see an example for early Venus in Extended Data Fig. 1, 325 W/m$^2$). This behaviour is visible through evaluations of the bond albedo (red and blue curves in Extended Data Fig. 3b) as well as the amount of reflected incoming solar radiation (red curves in Extended Data Fig. 2) which both substantially increase as the insolation approaches the water condensation insolation thresholds. This implies that a young and therefore hot terrestrial planet can form oceans on its surface only if the insolation that it receives is lower than the aforementioned insolation thresholds, as illustrated in Fig. 4 (red branches). States of stability where surface oceans are present (Fig. 4, blue branches) are possible only if (1) most water is initially condensed on the surface and (2) the insolation is below that of the runaway greenhouse[4,19]. The fact that the water condensation insolation threshold is low (owing to the nightside cloud feedback presented in this study) compared with the runaway greenhouse insolation threshold (which is higher, due to the dayside cloud feedback on Venus[4] and the dry descending branches of the Hadley cell on Earth[19]) produces a hysteresis loop with two possible stability states (Fig. 4), with water either fully condensed on the surface (blue branches) or fully evaporated (red branches) in the atmosphere.

These results have important consequences for our understanding of the comparative evolution of the atmosphere of Earth and Venus, as well as of rocky exoplanets. The minimal insolation received by Venus in the course of its evolution is about 500 W/m$^2$ (or about 1.47 times the Earth solar constant and about 0.75 times the Venus solar constant, around 4 billion years ago). As this minimal insolation is much higher than the water condensation insolation threshold (about 325 W/m$^2$), oceans should never have formed on the surface of Venus, which confirms and extends previous 1D cloud-free climate predictions[3]. Sensitivity studies show that this result is robust to a wide range of water and $CO_2$ contents (Methods). The water above the surface of Venus – always in the form of vapour – most probably gradually escaped from the atmosphere. It is not possible to tell from D/H measurements alone[35-37] whether this scenario is more likely than the one in which a late ocean was present[11]. However, recent estimates of water delivery[38] and atmospheric escape calculations[38,39] together suggest that water escaped very early from the atmosphere of Venus when the Sun's extreme ultraviolet radiation was stronger[40,41], in better agreement with our proposed scenario.

The insolation received on Earth today (about 340.5 W/m$^2$) is higher than the water condensation insolation threshold (about 312.5 W/m$^2$). It means first – as initially suggested with simplified 1D cloud-free climate models[28] – that the present-day Earth possesses three main stability states: (1) the present-day Earth with surface liquid oceans; (2) the snowball Earth[42] with oceans (quasi-) completely frozen; and (3) the steam Earth (this study), with water mostly in the form of steam. It also means that the reduced past solar luminosity (about 75% fainter 4 billion years ago) had a key role in allowing water to condense on Earth in the first place (by lowering the solar flux received on Earth below the water condensation insolation threshold) and form oceans, and thus for the emergence and development of life on Earth. This suggests that the faint young Sun, which has long been problematic in explaining the longevity of the Earth's liquid water oceans[21], may in fact have been a necessary opportunity for life to appear on Earth.

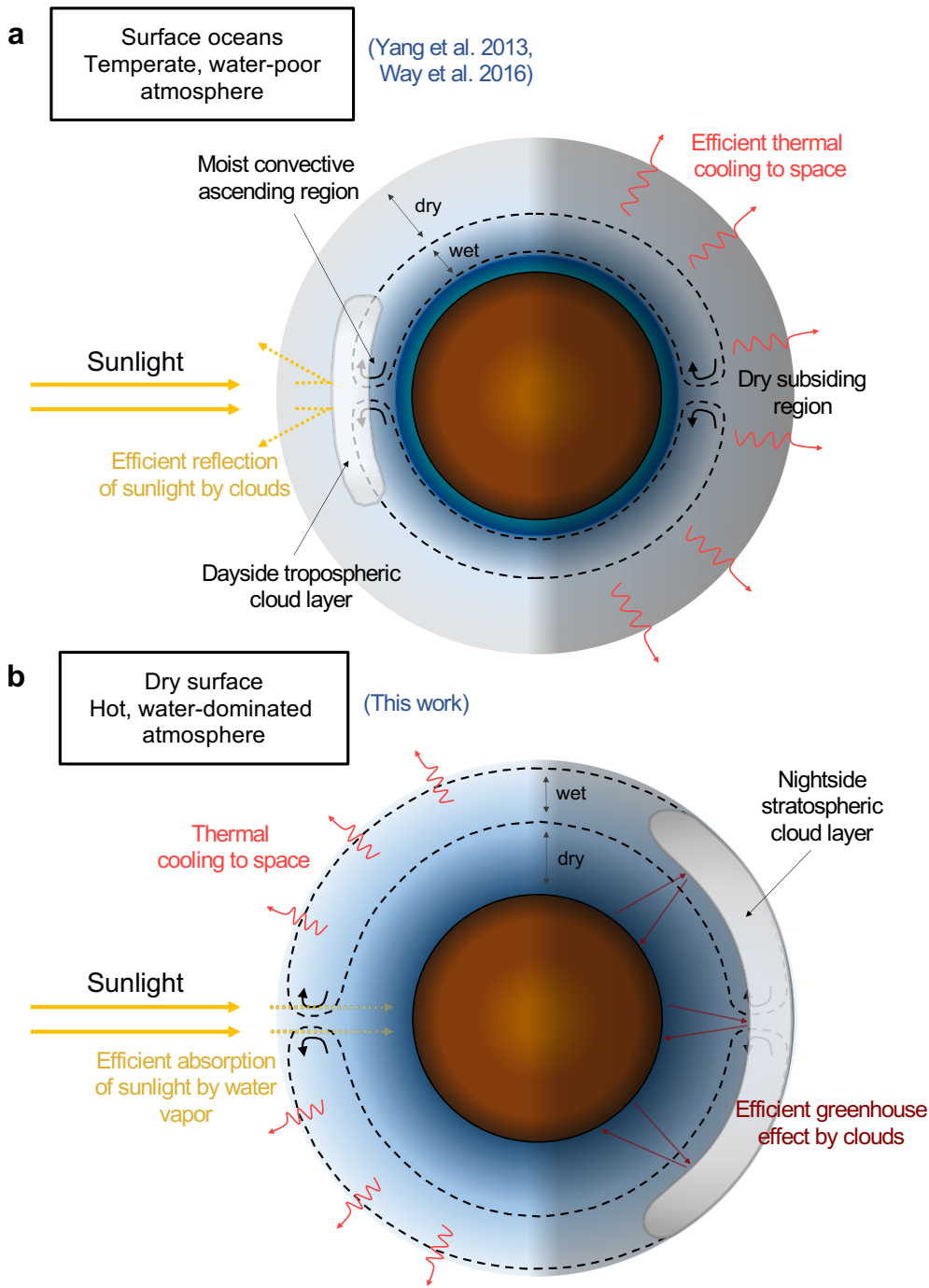

**Figure 1 | Day and night water cloud feedbacks. a,b,** Sketches of the cloud feedback mechanisms on slowly rotating planets initially covered with a liquid water ocean[4,31] **(a)** and fast and slowly rotating planets with oceans initially completely evaporated into the atmosphere (this study) **(b)**. Depending on their position (dayside or nightside), clouds have a very strong net cooling **(a)** or warming **(b)** effect.

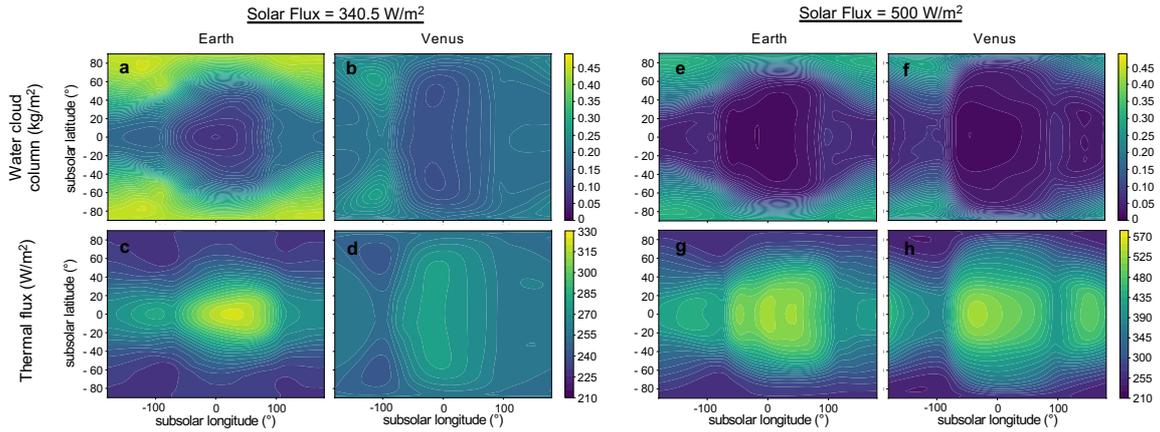

**Figure 2 | Water clouds and thermal emission horizontal maps. a-h** Maps of water cloud column (**a, b, e, f**) and thermal emission to space (**c, d, g, h**) for initially hot and steamy Earth ($P_{rot} \sim 24$ h; **a, c, e, g**) and Venus ($P_{rot} \sim 5{,}833$ h, that is, 244 times that of Earth; **b, d, f, h**), forced to insolations of 340.5 W/m² (that is, the present-day Earth insolation; **a-d**) and 500 W/m² (that is, the minimal insolation received on Venus, about 4 billion years ago when the Sun was 25% fainter than today; **e-h**). The maps were calculated in the heliocentric frame (that is, keeping the subsolar point at 0° longitude and 0° latitude), and using an average of five Earth years (for Earth simulations) and five Venusian days (for Venus simulations), respectively. In all simulations, clouds are preferentially located on the nightside, which markedly reduces the thermal emission to space.

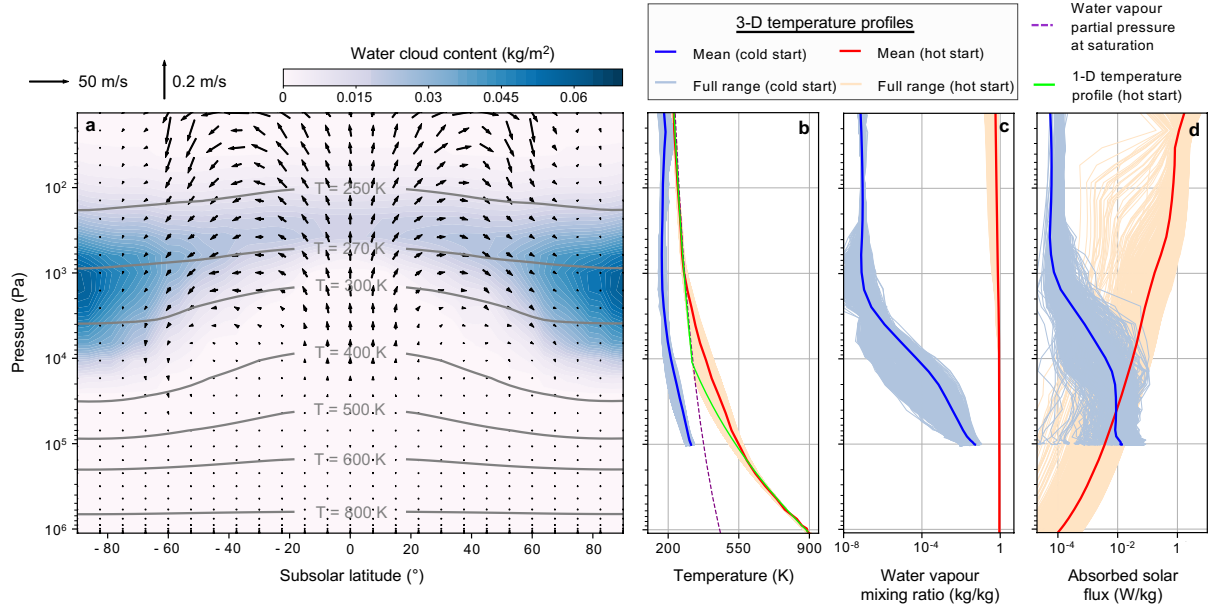

**Figure 3 | Mechanisms of nocturnal cloud formation. (a)** Annual-mean average zonal water cloud distribution (in the heliocentric frame) for initially hot and steamy Earth, forced at an insolation of 340.5 W/m$^2$, on which atmospheric temperature contours and local wind field have been superimposed. **(b-d)** Vertical profiles of the annual-mean globally averaged atmospheric temperatures **(b)**, water vapour mixing ratios **(c)** and absorbed solar flux **(d)**. The red vertical profiles were computed using the baseline simulation of hot and steamy Earth (340.5 W/m$^2$, P$_{H2O}$ = 10 bar: 'hot start'). The blue vertical profiles were computed for present-day Earth (aquaplanet) conditions, with all water initially condensed on the surface ('cold start'). The green temperature profile was computed using a 1D reverse radiative-convective model[29] of hot and steamy Earth as in refs[27,28]. The water vapour condensation curve (dotted purple) was calculated for the water vapour mixing ratio (0.91) of the warm and steamy simulations. Solar heating on the dayside pushes the tropopause downward so that clouds are formed by large scale radiative cooling of high-altitude air masses moving towards the nightside.

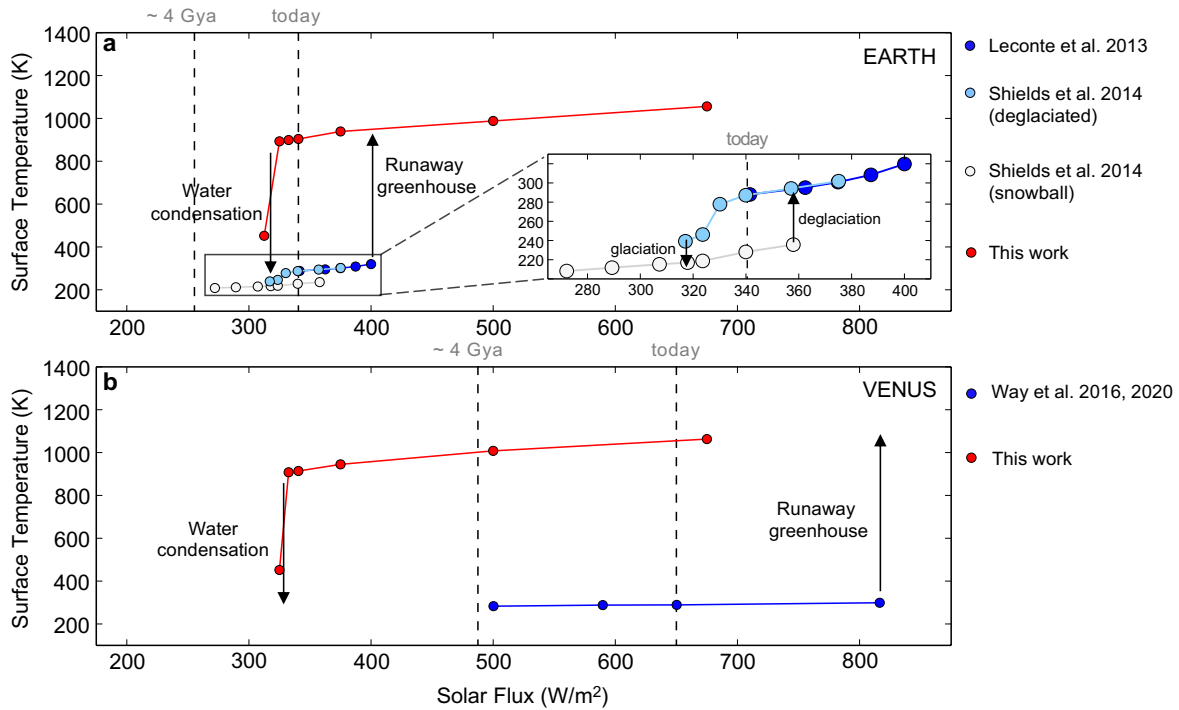

**Figure 4 | Hysteresis loops and conditions of ocean formation on Earth and Venus. a,b,** Surface temperature of Earth **(a)** and Venus **(b)** as a function of incoming solar flux. The blue branch corresponds to simulation results[4,11,19,43] that assume water is initially condensed on the surface. The red branch (this study) corresponds to initially hot and steamy simulation results, that assume water is initially in vapour form in the atmosphere. For reference, insolation is 340.5 W/m$^2$ on present-day Earth and is 650 W/m$^2$ on present-day Venus. Insolations were 25% lower about 4 billion years ago (Ga). Although the insolation on Earth 4 billion years ago was low enough for it to reach the blue branch and thus form surface oceans, the insolation on initially hot and steamy Venus forced it directly into the red branch, preventing surface water condensation.

## METHODS

**1. Numerical climate model and simulations setup**

The simulations were performed with the LMD Generic Global Climate Model (GCM). The model, which was historically developed to study the climate of the Earth[44] and then Mars[45], has been upgraded to simulate solar system planet paleoclimates[17,46–49] and exoplanet climates[50–52]. The model uses the 3D dynamical core LMDz, which is based on a finite-difference formulation of the primitive equations of geophysical fluid dynamics. The GCM includes an up-to-date generalized radiative transfer (described in 'Radiative transfer in $H_2O$-rich atmospheres') that takes into account the absorption and scattering by the atmosphere, the clouds and the surface. The GCM includes condensation, evaporation, sublimation, and precipitation of water ($H_2O$) (described in 'Turbulence, convection and clouds in $H_2O$-rich atmosphere'). For the present study, special care has been taken (see hereafter) to treat properly the situations in which the atmosphere is hot, thick and water-vapour-dominated. More details on the model can be found in refs[19,50,52] (and references therein).

Our baseline simulations were computed at a horizontal resolution of 64 × 48 grid points (in longitude × latitude). In the vertical direction, the model is composed of 34 distinct atmospheric layers (using σ hybrid coordinates, where σ is the ratio between pressure and surface pressure) for the baseline simulations (between 31 and 40 for sensitivity studies), ranging from the surface up to ∼ 10 Pa. The dynamical (for example, to compute winds) time step of the simulations is about 40 s; the physical (for example, to compute convection) time step is about 6 min; the radiative (for example, to compute radiative transfer) time step is about 1 h.

Simulation input parameters for our baseline and sensitivity GCM simulations are summarized in Extended Data Table 1. The initial state of the simulations and the convergence method are described in 'A numerical scheme to accelerate the convergence of simulations'.

**2. Radiative Transfer in $H_2O$-rich atmospheres**

Our climate model includes a generalized radiative transfer code based on the correlated-k method[53] suited for fast calculations. We adapted the radiative transfer to any mixture of nitrogen ($N_2$) and $H_2O$ gases (as well as $CO_2$ and $H_2O$ gases for sensitivity studies, as in ref[49]).

Our radiative transfer calculations are performed on 55 spectral bands in the thermal infrared (from 0.65 to 100 µm) and 45 in the visible domain (from 0.3 to 6.5 µm). The infrared channel has been extended to relatively short wavelengths (0.65 µm) to capture the thermal emission of the lowest and warmest (more than about $10^3$ K in some simulations) layers. Sixteen non-regularly spaced grid points were used for the *g*-space integration, where *g* is the cumulative distribution function of the absorption for each band. $N_2$-$N_2$[54], $H_2O$-$H_2O$[55] and $H_2O$-$N_2$[55] (using MT CKD version 3.2, from 200 K to 1,500 K and from 0 to 20,000 cm$^{-1}$) continua were accounted for. Correlated-k tables were computed as in ref[19] for $N_2$ and $H_2O$ (and as in ref[49] for $CO_2$ and $H_2O$, in sensitivity studies), using the HITRAN 2008[56] database for wavelength above 1µm and the HITEMP 2010[57] database below 1µm, following the results of ref[28].

We performed 1D cloud-free radiative-convective model calculations[29] using the radiative transfer parameterizations described above and found that the planetary albedo and thermal emission asymptotic limits calculations match very closely that of state-of-the-art 1D calculations of refs[27,28] (see Extended Data Fig. 3). This validates the choices made for our radiative transfer calculations, which is used as is in the 3D simulations presented in this study.

The exact temperature at which the thermal emission departs from the asymptote depends very strongly on the water content in the atmosphere[29]. This is why we only compare here the values of the asymptote limit of refs[27,28] which have been shown to be weakly dependent on the total water content[29].

**3. Turbulence, convection and clouds in $H_2O$-rich atmospheres**

Subgrid-scale dynamical processes (turbulent mixing and convection) were parameterized as in ref[19]. The planetary boundary layer was accounted for by the time-dependent 2.5-level closure scheme[58,59], and complemented by a convective adjustment which rapidly mixes the atmosphere in the case of unstable temperature profiles. Moist convection was taken into account following a moist convective adjustment scheme that originally derives from the 'Manabe scheme'[47,60]. In the version of our scheme, relative humidity is let free and limited to 100% (supersaturation is not permitted). In practice, when an atmospheric grid cell reaches 100% saturation and the corresponding atmospheric column has an unstable temperature vertical profile, the moist convective adjustment scheme is performed to get a stable moist adiabatic lapse rate. We used a generalized formulation of the moist-adiabat lapse rate developed in ref[19] to account for the fact that water vapour is both the dominant and the condensible gas in our simulations. We also used the numerical scheme proposed in ref[19] to account for atmospheric mass change after the condensation or the evaporation of water vapour.

When condensing, water vapour form liquid water droplets or water ice particles, depending on the atmospheric temperature and pressure[49,61], forming clouds. We used a fixed number of activated cloud condensation nuclei (CCNs) per unit mass of air $N_c$ to determine the local H$_2$O cloud particle sizes, based on the amount of condensed material. $N_c$ was taken to be constant everywhere in the atmosphere, and equal to $10^5$ kg$^{-1}$ for both liquid and ice water clouds in the baseline simulations (but varied in sensitivity simulations). This value roughly corresponds to that of high-altitude clouds on Earth[62] and is also a standard value used for studies of rocky planet climates[47,51,63].

The effective radius $r_{eff}$ of the cloud particles is then given by:

$$r_{eff} = \left(\frac{3\, q_c}{4\, \pi\, \rho_c\, N_c}\right)^{1/3}$$

where $\rho_c$ is the density of the cloud particles (1,000 kg/m$^3$ for liquid and 920 kg/m$^3$ for water ice) and $q_c$ is the mass mixing ratio of cloud particles (in kg per kg of air). The effective radius of the cloud particles is then used to compute (1) their radiative properties calculated by Mie scattering and (2) their sedimentation velocity. Water precipitation is divided into rainfall and snowfall, depending on the nature (and thus the temperature) of the cloud particles. Rainfall is parameterised using the scheme of ref[64], accounting for the conversion of cloud liquid droplets to raindrops by coalescence with other droplets. Rainfall is considered to be instantaneous (that is, it goes directly to the surface) but can evaporate while falling through subsaturated layers. The evaporation rate of precipitation $E_{precip}$ (in kg/m$^3$/s) is determined by[65]:

$$E_{precip} = 2 \times 10^{-5} \left(1 - \frac{q_v}{q_{s,v}}\right) \sqrt{F_{precip}}$$

where $q_v$ and $q_{s,v}$ are the water vapour mixing ratios in the air cell and at saturation, respectively. $F_{precip}$ is the precipitation flux (in kg/m$^2$/s). In hot and steamy simulations, re-evaporation of precipitation is always complete, that is, precipitation always fully evaporate in the dry lower atmosphere before it reaches the ground. The snowfall rate is calculated using the sedimentation velocity of particles $V_{sedim}$ (in m/s), assumed to be equal to the terminal velocity that we approximate by a Stokes law, with a 'slip-flow' correction factor[66].

**4. A numerical scheme to accelerate the convergence of simulations**

The very high opacity and heat capacity of multi-bar water-dominated atmospheres considerably increases the computing time required for simulations to reach top-of-atmosphere radiative equilibrium. To overcome this issue, we implemented a numerical scheme designed to accelerate the convergence of GCM simulations towards equilibrium.

We start first with the GCM simulations at highest insolations (675 and 500 W/m²) with a partial pressure of $H_2O$ and $N_2$ of 10 bar and 1 bar (in the baseline simulations), respectively, and an isothermal atmospheric temperature profile at 1,050 K. Although the upper atmospheric layers evolve very rapidly towards equilibrium, we had to implement a new strategy to converge the deepest layers of the atmosphere. Following ref[63], we artificially multiplied the heating rates (in shortwave and longwave channels) of the radiative transfer by a factor $\left(\frac{P}{P_{lim}}\right)^\alpha$ for atmospheric pressures $P > P_{lim}$, with $P_{lim} = 10^4$ Pa and α a convergence factor. As the simulations get closer to convergence, we decrease the value of α from 0.5 to 0.3 and eventually to 0. We consider that the simulations have reached their final equilibrium state when the top-of-the-atmosphere radiative imbalance is typically lower than about 1 W/m².

We repeated the same procedure for simulations at lower insolations, except that we used the final equilibrium state of simulations at higher insolations as the initial state (see Extended Data Fig. 1, which shows the temporal evolution of the hot and steamy Venus simulations at several insolations).

On the whole, and taking advantage of the convergence scheme presented above, we have managed to converge all the simulations presented in this work using a total of around 600 kh CPU (central processing unit) on the French supercomputer OCCIGEN. Calculating the threshold insolation at which water condenses on the surface (which requires running simulations at multiple insolations) required around 150 kh CPU for each configuration explored.

To check that our accelerated convergence scheme is working effectively, we have performed several convergence tests (for the simulation of hot and steamy Venus, at an insolation of 500 W/m$^2$). We performed first a simulation in which we started with an isothermal atmospheric profile at 700 K, and the simulation converged (less quickly though) towards the same final equilibrium. We have also performed two GCM simulations in which we started with a cold adiabatic profile ($T_{surf}$ = 700 K) and a hot adiabatic profile ($T_{surf}$ = 1,400 K) calculated with a 1D reverse numerical model[29]. Again, the simulations converged towards the same final equilibrium. Eventually, we have performed a last simulation in which we tested an alternative convergence scheme. After starting from an isothermal atmospheric profile at 1,050 K, we proceeded with the following repeated steps:

1. We run the GCM for two Venusian days.

2. We extrapolate the evolution of the temperature field at time $t$ over $n_{days}$ (the number of Venusian days) using:

$$T_{i,j,k}(t + n_{days}) = T_{i,j,k}(t) + n_{days} \times \Delta T_{mean,k}$$

with $T_{i,j,k}$ the temperature at the $i,j,k$ spatial coordinates (corresponding respectively to longitude, latitude and altitude coordinates) and $\Delta T_{mean,k}$ the change (over the second Venusian day of GCM simulations) of the mean horizontal (averaged over all longitudes and latitudes) temperature field at the altitude layer $k$. $n_{days}$ was first set arbitrarily to 50, then 10 and eventually 0, when the planetary atmosphere is close to convergence. We limit the variations of temperature to 50 K to avoid numerical instabilities.

3. We repeat the previous steps, until the top-of-the-atmosphere radiative imbalance is lower than about 1 W/m$^2$.

With this alternative convergence scheme, the simulation converged (less quickly though) towards the same final equilibrium. These sensitivity tests validate the use of our accelerated convergence scheme.

## 5. Comparisons with previous 1D modelling

We first validated our radiative transfer parameterizations and calculations by comparing the results of our cloud-free 1D numerical climate calculations[29] with that of refs[27,28]. Extended Data Figure 3 shows that the moist tropospheric asymptotic limit – sometimes also known as the Simpson- Nakajima limit[67–69] – matches closely that of refs[27,28].

Extended Data Figure 3 also shows comparisons with the results of our baseline 3D GCM simulations of hot and steamy Earth and Venus. Differences with 1D calculations arise from the effects of dynamics, convection, radiation and mostly clouds. The dayside cloud coverage decreases with increasing insolation (and thus, the surface temperature), which is reflected in the evolution of the bond albedo with surface temperature. For the highest insolation explored in this study (about 675 W/m$^2$, corresponding to a surface temperature of ~ 1,060 K for Earth and Venus, in Extended Data Fig. 3b), the bond albedo of 3D simulations is very close to that predicted with state-of-the-art 1D cloud-free calculations[27–29].

Thermal emission to space is significantly different from that predicted with 1D cloud-free calculations for two reasons. First, nightside clouds are always present for the range of insolations (~ 325-675 W/m$^2$) explored in this study. These clouds produce a strong greenhouse effect that significantly reduces thermal emission (Extended Data Fig. 2). This is well visible at low insolation (or at the lowest surface temperatures) in Extended Data Fig. 3a. Second, the 1D reverse calculations[27–29] assume fixed temperature profiles (that is, with fixed stratospheric temperatures, at 200 K, and convective profiles otherwise; this is illustrated by the green profile in Fig. 3b). Our 3D baseline calculations do take into account the direct warming of the atmosphere by the incoming solar flux, which produces significant heating at the subsolar point (see Fig. 3d), and thus higher stratospheric temperatures at the subsolar point (see Fig. 3b), which produces a rapid increase (compared with 1D, reverse models) of the thermal emission to space with insolation (see Extended Data Fig. 3a). This effect could be implemented in 1D reverse numerical climate models by implementing a parameterisation of the vertical temperature structure as a function of the insolation.

## 6. Sensitivity studies

We proceeded to several sensitivity studies to test the robustness of the nightside cloud feedback warming mechanism.

**i. varying cloud microphysical properties**

We performed a first sensitivity study (for hot and steamy Venus, insolation at 500 W/m$^2$, that is, that of Venus about 4 billion years ago) in which we varied the number of Cloud Condensation Nuclei (CCN), the key parameter[46,63] of the clouds microphysics model used in our GCM, for a wide range of values from 10$^3$ to 10$^7$ per kg of atmosphere. The results, which are summarized in Extended Data Fig. 4, reveal that the dichotomic structure of the clouds (absence of clouds on the dayside, presence of clouds on the nightside) and the net warming effect of clouds are robust to the choice of the number of CCNs. The net radiative effect of clouds increases with the number of CCNs. It is equal to 20, 58, 98, 104 and 108 W/m$^2$ for a CCN number of 10$^3$, 10$^4$, 10$^5$, 10$^6$, 10$^7$ kg$^{-1}$, respectively.

**ii. varying the amount of water**

Our baseline 3D GCM simulations of hot and steamy Earth and Venus were computed assuming a $H_2O$ partial pressure of 10 bar, corresponding to around 100 m GEL on Earth and 110 m GEL on Venus. We performed a sensitivity study (for hot and steamy Venus, insolation at 500 W/m$^2$) in which we varied the $H_2O$ partial pressure between 1 bar and 30 bar, corresponding to around 11-330 m GEL on Venus. Simulations at 30 bar of $H_2O$ show very similar behaviour to those at 10 bar, which results from the fact that the pressure is high enough to decouple the surface from the upper part of the atmosphere. We therefore expect that simulations with a partial pressure of $H_2O$ larger than 30 bar (that is, GEL larger than around 300 m) also show a similar behaviour. This means our simulations are roughly representative of the range of past water content estimates on Venus[4].

Extended Data Fig. 5 summarizes the results of this sensitivity study. The net warming effect of clouds is similar across simulations. We also observe, as predicted in 1D numerical climate models[29], that the surface temperature increases with $H_2O$ partial pressure. The bond albedo and thermal emission are relatively stable across simulations.

**iii. adding carbon dioxide**

Our baseline 3D GCM simulations of hot and steamy Earth and Venus were computed assuming a $N_2$ partial pressure of 1 bar.

We have performed two sensitivity experiments in which $N_2$ was replaced by 1 bar and 10 bar of $CO_2$. We carried out these experiments to take into account the possibility that $CO_2$ was present in large quantities in the atmosphere of early Venus[70] as it is today. Here again, we observe that clouds preferentially form on the nightside, having thus a net warming effect (see Extended Data Fig. 5).

We observe that the bond albedo increases with $CO_2$ partial pressure, as predicted by previous $CO_2+H_2O$ 1D modeling[71], due to $CO_2$ having a larger Rayleigh scattering cross-section than $H_2O$. The bond albedo is lower than that of 1D cloud-free $CO_2+H_2O$ simulations of ref[71], not only because the 3D GCM simulations are depleted in dayside clouds, but also because some visible water lines – present in HITEMP[57] and not in HITRAN[56], which was used in ref[71] – increase the absorption, further lowering the albedo[28].

**iv. influence of the initial cloud cover distribution**

We performed a last sensitivity study (for hot and steamy Venus, insolation at 500 W/m$^2$) in which we artificially forced the cloud cover to be complete on both the night and day sides, in order to explore whether clouds on the dayside could push the planet towards cooling and thus ultimately surface water condensation. The sensitivity simulation shows that the cloud distribution very quickly regains a dichotomous structure after a few Earth days of simulation only.

**ACKNOWLEDGEMENTS**

This project has received funding from the European Union's Horizon 2020 research and innovation program under the Marie Sklodowska-Curie Grant Agreement No. 832738/ESCAPE. This project has received funding from the European Research Council (ERC) under the European Union's Horizon 2020 research and innovation program (grant agreements No. 724427/FOUR ACES and 679030/WHIPLASH). This work has been carried out within the framework of the National Centre of Competence in Research PlanetS supported by the Swiss National Science Foundation. We acknowledge the financial support of the SNSF. M.T. thanks the Gruber Foundation for its support to this research. M.T. thanks N. Chaniaud for her help in preparing Fig. 1. We thank the LMD Generic global climate model team for the

**DATA AVAILABILITY**

The data that support the findings of this study are available at https://doi.org/10.5281/zenodo.4680905. Source data are provided with this paper.

**CODE AVAILABILITY**

The LMD Generic global climate model code (and documentation on how to use the model) used in this work can be downloaded from the SVN repository https://svn.lmd.jussieu.fr/Planeto/trunk/LMDZ.GENERIC/ (version 2528). More information and documentation are available on http://www-planets.lmd.jussieu.fr.

**AUTHOR CONTRIBUTIONS**

M.T. developed the core ideas of the manuscript, developed and performed the 3D GCM simulations, wrote the manuscript and prepared the figures. E.B. and G.C. provided advice on sensitivity studies. D.E., G.C. and J.L. provided advice on the structure of the figures. J.L. provided advice on the organization of the manuscript, as well as for understanding the mechanism of cloud formation. E.M. provided advice on literature selection. All authors provided guidance and comments on the manuscript.

## COMPETING INTERESTS

The authors declare that they have no competing financial interests.

## CORRESPONDENCE

Correspondence and requests for materials should be addressed to M.T. (martin.turbet@unige.ch).

| Physical Parameters | Values (Earth) | Values (Venus) |
| --- | --- | --- |
| Solar Flux (W/m$^2$) | 312.5, 325, 332.5, 340.5, 375, 500, 675 | |
| Obliquity (°) | 23.5 | 177 |
| Orbital eccentricity | 0 | |
| Rotation period (h) | 24 | 5,833 |
| Solar day (h) | 24 | 2,802 |
| Radius (km) | 6,370 | 6,052 |
| Gravity (m/s$^2$) | 9.81 | 8.87 |
| Bare ground albedo | 0.2 | |
| Ground thermal inertia (J/m$^2$/s$^{1/2}$/K) | 2,000 | |
| Surface topography | flat | |
| Surface roughness coefficient (m) | 0.01 | |
| H$_2$O partial pressure (bar) | 10 | 10 * |
| N$_2$ partial pressure (bar) | 1 | 1 † |
| CO$_2$ partial pressure (bar) | 0 | 0 ‡ |
| Number of H$_2$O cloud condensation nuclei (CCN) for ice and liquid (kg$^{-1}$) | 10$^5$ | 10$^5$ § |

**Extended Data Table 1 | Summary of the main physical parameters used in the GCM for the baseline hot and steamy Earth and Venus simulations.** *We varied H$_2$O partial pressure from 1 to 30 bar in some hot and steamy Venus GCM sensitivity simulations; †We fixed N$_2$ partial pressure to 0 bar in hot and steamy Venus GCM sensitivity simulations with CO$_2$; ‡In hot and steamy Venus GCM sensitivity simulations with CO$_2$, we used partial pressure of 1 and 10 bar of CO$_2$; §We varied the number of CCN from 10$^3$ to 10$^7$ in GCM sensitivity simulations of Venus.

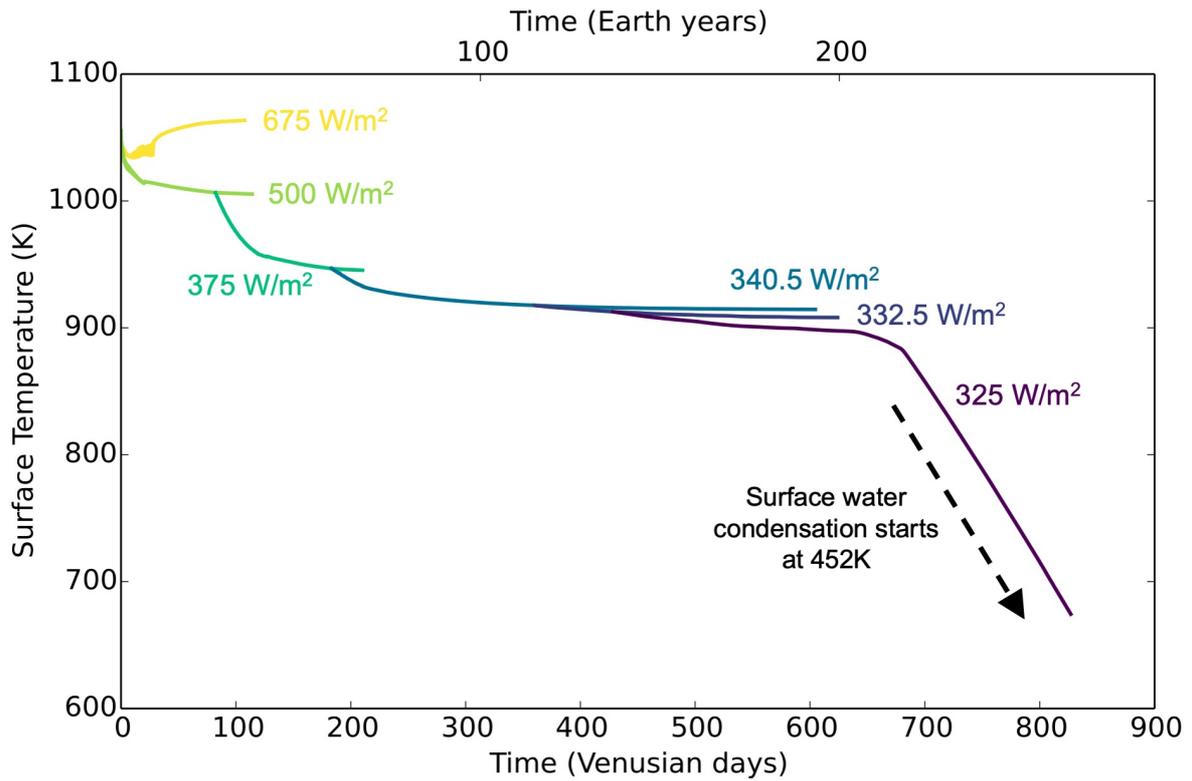

**Extended Data Figure 1 | Temporal evolution of modelled surface temperatures.**
Temporal evolution of the globally-averaged surface temperatures in the 3D GCM baseline simulations of (initially hot and steamy) Venus, for several insolations.

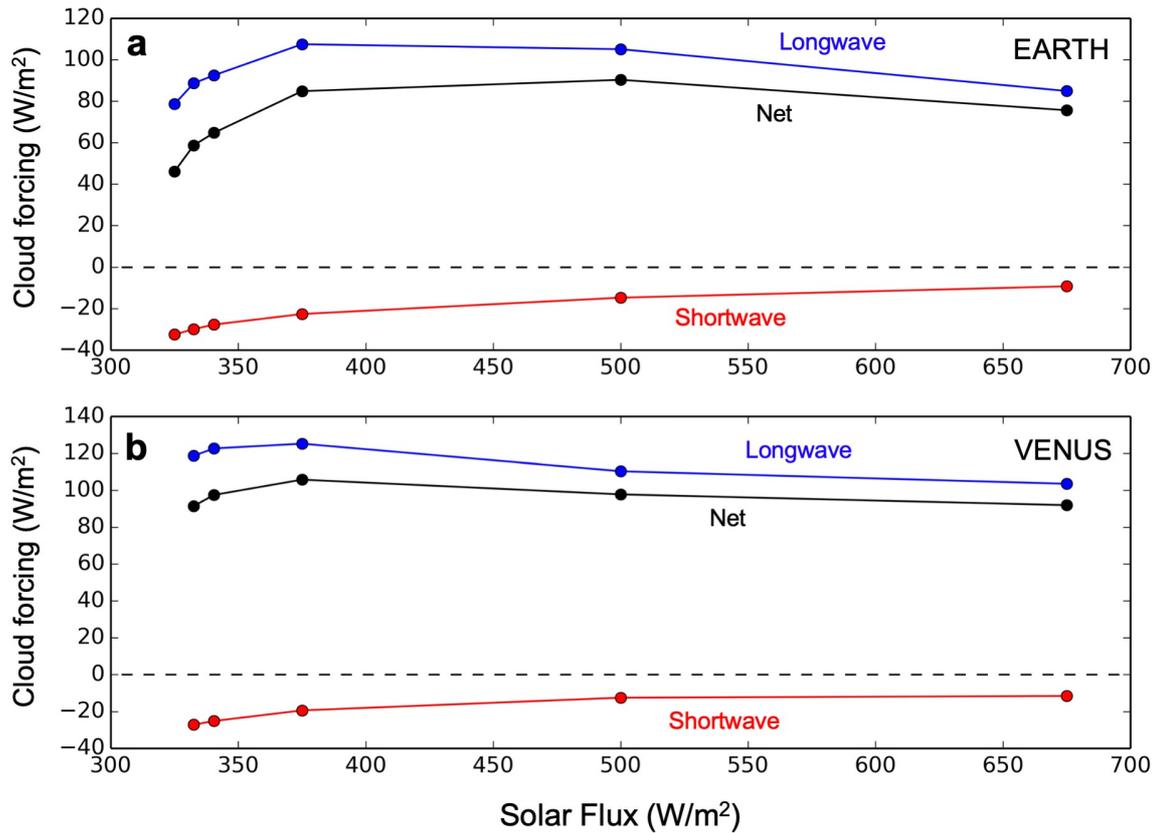

**Extended Data Figure 2 | Cloud forcings in hot and steamy early Venus and Earth simulations.** Radiative balance of clouds on hot and steamy Earth **(a)** and Venus **(b)** as a function of the incident solar flux. Blue curves indicate the greenhouse effect of clouds. Red curves indicate the amount of incoming solar radiation reflected back by the clouds (the more negative the value, the greater the reflected flux.). Black curves indicate the net radiative effect of clouds (positive values mean warming). In all initially hot and steamy simulations, clouds lead to a strong atmospheric warming.

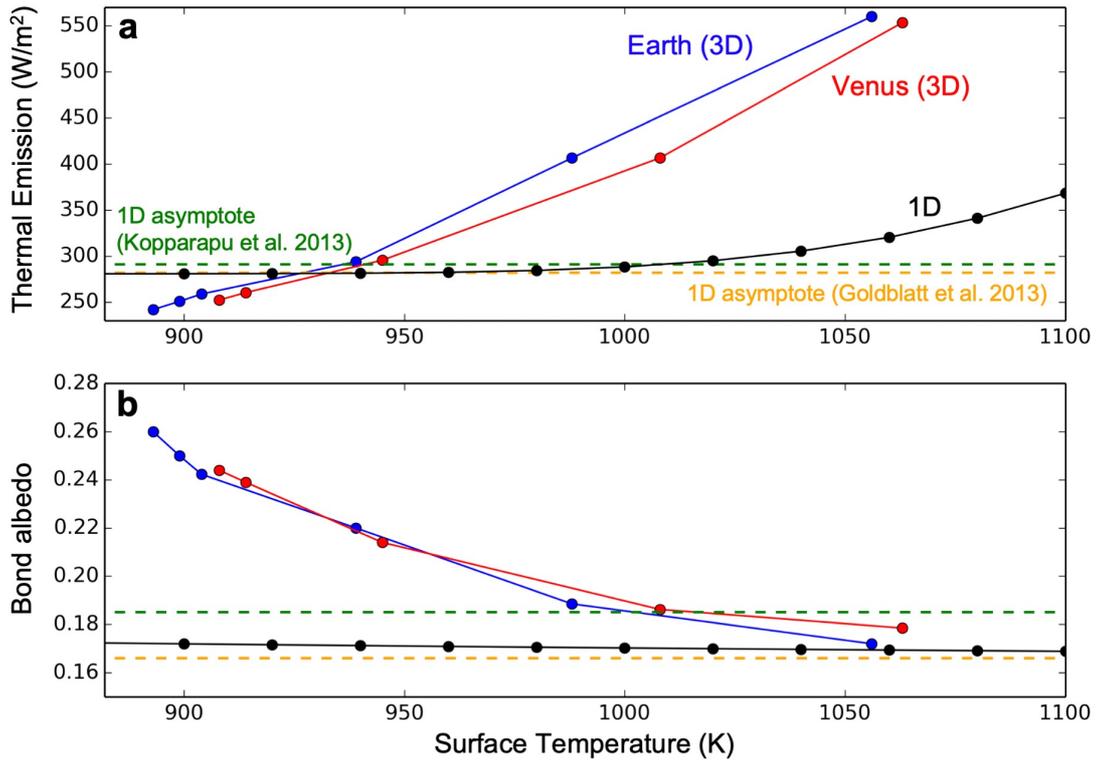

**Extended Data Figure 3 | Radiative budget comparisons between 1D and 3D models.**
Thermal emission to space (a) and bond albedo (b) as a function of the surface temperature for our 3D GCM simulations of Earth (blue) and Venus (red). We have also added the results of 1D radiative-convective cloud-free calculations[29] (in black), using $H_2O$ and $N_2$ partial pressures of 10 and 1 bar, respectively, as in the 3D baseline simulations. For comparison, we added the moist tropospheric radiation limits[68] for the thermal emission[67] and the bond albedo from refs[27,28]

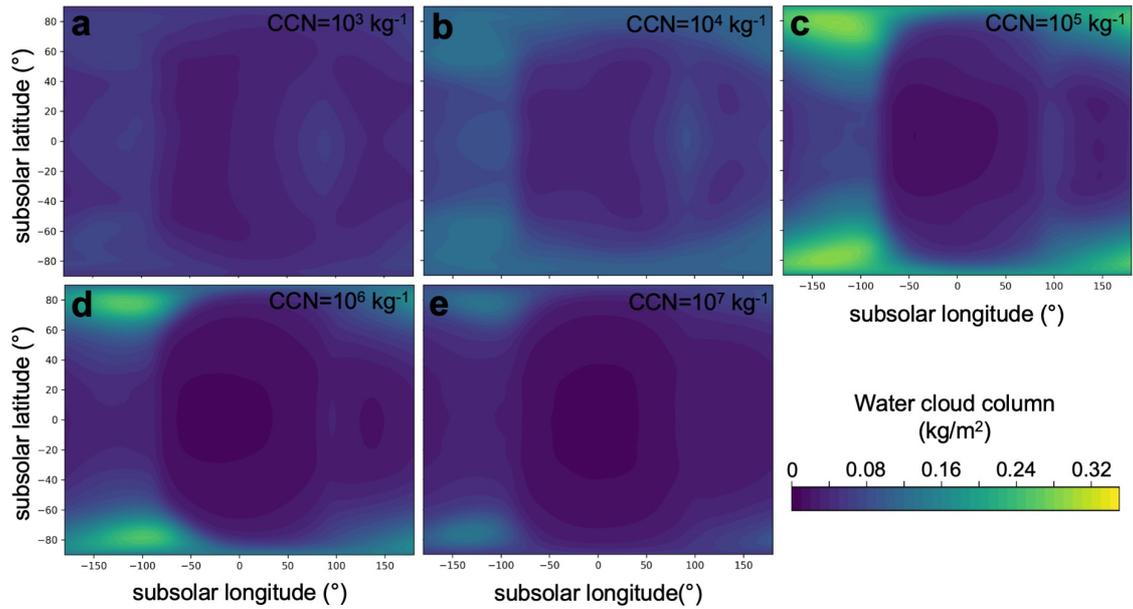

**Extended Data Fig. 4 | Impact of cloud microphysical properties on their spatial distribution.** Maps of water cloud column for early Venus (at a top-of-atmosphere insolation of 500 W/m$^2$, i.e. the minimal insolation received on Venus, about 4 billion years ago when the Sun was 25% fainter than today), with different cloud microphysics parameterisations ($10^3$, $10^4$, $10^5$, $10^6$ and $10^7$ Cloud Condensation Nuclei (CCN) per kg of air, for panels a, b, c, d and e, respectively). The maps were calculated in the heliocentric frame (i.e., keeping the subsolar point at 0° longitude and 0° latitude), and using an average of two Venusian days. The distribution of clouds (present on the nightside, absent on the dayside) is robust to the choice of the number of CCN.

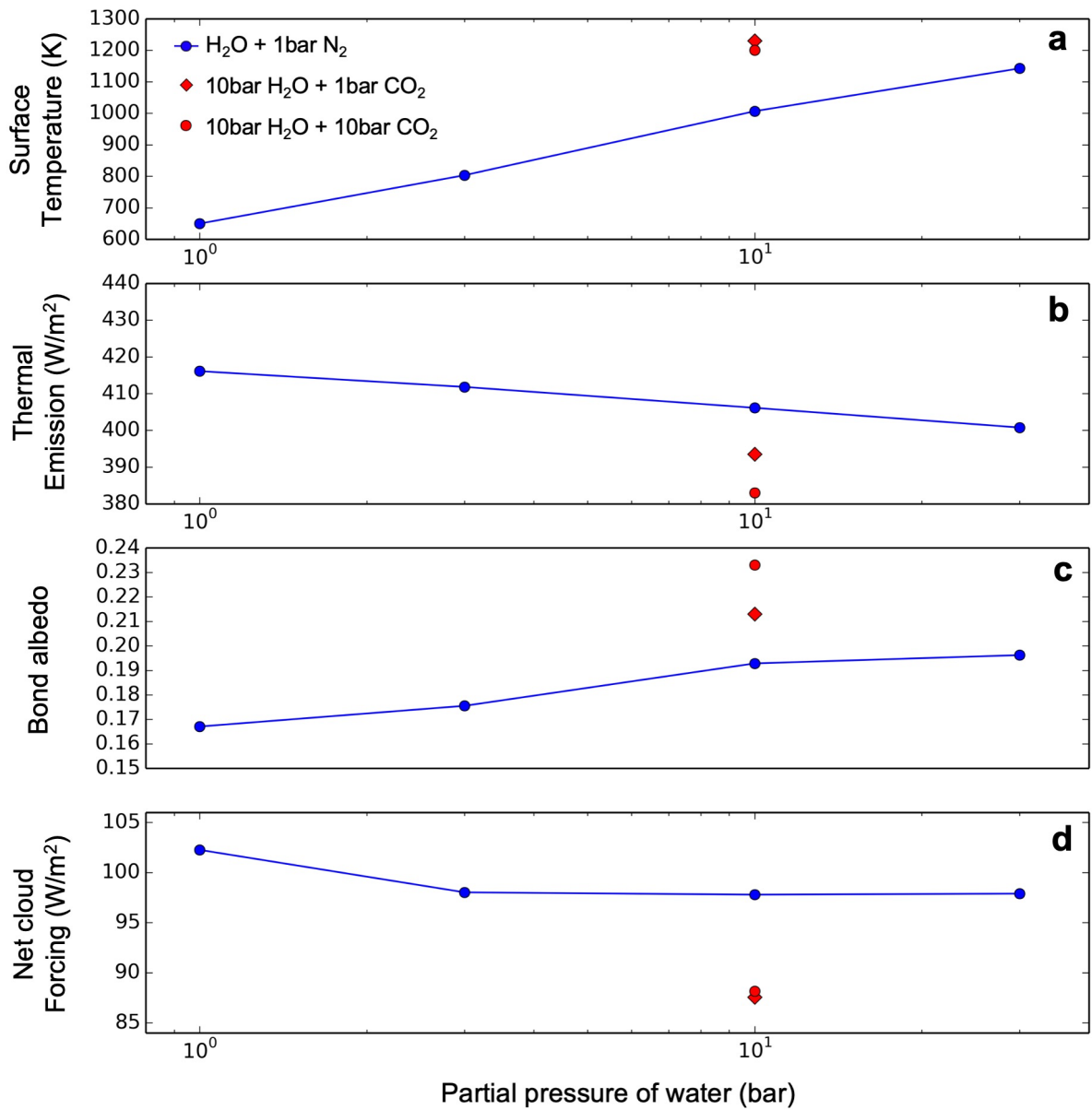

**Extended Data Fig. 5 | Effects of water and carbon dioxide atmospheric contents.** Impact of the water and carbon dioxide atmospheric contents on the surface temperature (a), thermal emission to space (b), bond albedo (c) and net cloud radiative forcing (d). The calculations assume a hot and steamy Venus (insolation at 500 W/m$^2$) with 1 bar of $N_2$ and between 1 and 30 bar of $H_2O$ (in blue); with 1-10 bar of $CO_2$ and 10 bar of $H_2O$ (in red).